# A Secure and High Capacity Image Steganography Technique

Hemalatha S[1], U Dinesh Acharya[2], Renuka A[3], Priya R. Kamath[4]

[1,2,3,4]Department of Computer Science and Engineering, Manipal Institute of Technology, Manipal University, Manipal, Karnataka, India
[1]hema.shama@manipal.edu; [2]dinesh.acharya@manipal.edu; [3]renuka.prabhu@manipal.edu; [4]priyarkamath@gmail.com

## ABSTRACT

*Steganography is the science of "invisible" communication. The purpose of Steganography is to maintain secret communication between two parties. The secret information can be concealed in content such as image, audio, or video. This paper provides a novel image steganography technique to hide multiple secret images and keys in color cover image using Integer Wavelet Transform (IWT). There is no visual difference between the stego image and the cover image. The extracted secret images are also similar to the original secret images. Very good PSNR (Peak Signal to Noise Ratio) values are obtained for both stego and extracted secret images. The results are compared with the results of other techniques, where single image is hidden and it is found that the proposed technique is simple and gives better PSNR values than others.*

## KEYWORDS

*Steganography, IWT, MSE, PSNR, RGB, Luminance, Chrominance*

## 1. INTRODUCTION

Information security is essential for confidential data transfer. Steganography is one of the ways used for secure transmission of confidential information. It contains two main branches: digital watermarking and steganography. The former is mainly used for copyright protection of electronic products while, the latter is a way of covert communication. Avoiding communication through well-known channels greatly reduces the risk of information being leaked in transit. Hiding information in a photograph of the company picnic is less suspicious than communicating an encrypted file.

The main purpose of steganography is to convey the information secretly by concealing the very existence of information in some other medium such as image, audio or video. The content used to embed information is called as cover object. The cover along with the hidden information is called as stego-object [1]. In this paper color image is taken as cover and two grey scale images are considered as secret information. Secret images and stego keys are embedded in the cover image to get stego image. The major objective of steganography is to prevent some unintended observer from stealing or destroying the confidential information. There are some factors to be considered when designing a steganography system: [1]

- Invisibility: Invisibility is the ability to be unnoticed by the human.
- Security: Even if an attacker realizes the existence of the information in the stego object it should be impossible for the attacker to detect the information. The closer the stego image to the cover image, the higher the security. It is measured in terms of PSNR.

$$\text{PSNR} = 10 \log \frac{L^2}{\sqrt{MSE}} \text{dB} \qquad (1)$$

where L = maximum value, MSE = Mean Square Error.

$$MSE = \frac{1}{N} \sum_{i=1}^{N} |Xi - Xi'|^2 \quad (2)$$

where X = original value, X' = stego value and N = number of samples.

High PSNR value indicates high security because it indicates minimum difference between the original and stego values. So no one can suspect the hidden information.

- Capacity: The amount of information that can be hidden relative to the size of the cover object without deteriorating the quality of the cover object
- Robustness: It is the ability of the stego to withstand manipulations such as filtering, cropping, rotation, compression etc.

The design of a steganographic system can be categorized into spatial domain methods and transform domain methods [1]. In spatial domain methods, the processing is applied on the image pixel values directly. The advantage of these methods is simplicity. The disadvantage is low ability to bear signal processing operations. Least Significant Bit Insertion methods, Pallet based methods come under this category. In transform domain methods, the first step is to transform the cover image into different domain. Then the transformed coefficients are processed to hide the secret information. These changed coefficients are transformed back into spatial domain to get stego image. The advantage of transform domain methods is the high ability to face signal processing operations. However, methods of this type are computationally complex. Steganography methods using DCT (Discrete Cosine Transforms), DWT (Discrete Wavelet Transforms), IWT, DFT (Discrete Fourier Transforms) come under this category.

In this paper the secret images are embedded using IWT. The Wavelet Transform provides a time-frequency representation of the signal. IWT is a more efficient approach to lossless compression. The coefficients in this transform are represented by finite precision numbers which allows for lossless encoding. This wavelet transform maps integers to integers. In case of Discrete Wavelet Transform, if the input consists of integers (as in the case of images), the resulting output no longer consists of integers. Thus the perfect reconstruction of the original image becomes difficult. However, with the introduction of Wavelet transforms that map integers to integers the output can be completely characterized with integers. The LL sub-band in the case of IWT appears to be a close copy with smaller scale of the original image while in the case of DWT the resulting LL sub-band is distorted slightly, as shown in Figure 1.[2].

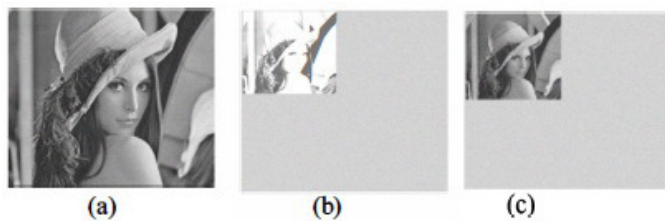

Figure 1. (a) Original image Lena. (b) One level DWT in sub band LL (c) One level IWT in sub-band LL.

If the original image (I) is X pixels high and Y pixels wide, the level of each of the pixel at (i,j) is denoted by $I_{i,j}$.[3]

The IWT coefficients are given by

$$LL_{i,j} = \lfloor (I_{2i, 2j} + I_{2i+1, 2j})/2 \rfloor \quad (1)$$
$$HL_{i,j} = I_{2i+1, 2j} - I_{2i, 2j} \quad (2)$$

$$LH_{i,j} = I_{2i, 2j+1} - I_{2i, 2j} \qquad (3)$$

$$HH_{i,j} = I_{2i+1, 2j+1} - I_{2i, 2j} \qquad (4)$$

The inverse transform is given by

$$I_{2i, 2j} = LL_{i,j} - \lfloor HL_{i,j}/2 \rfloor \qquad (5)$$

$$I_{2i, 2j+1} = LL_{i,j} + \lfloor (HL_{i,j+1})/2 \rfloor \qquad (6)$$

$$I_{2i+1, 2j} = I_{2i, 2j+1} + LH_{i,j} - HL_{i,j} \qquad (7)$$

$$I_{2i+1, 2j+1} = I_{2i+1,2j} + HH_{i,j} - LH_{i,j} \qquad (8)$$

where, $1 \leq i \leq X/2$, $1 \leq j \leq Y/2$ and $\lfloor \rfloor$ denotes floor value.

## 2. RELATED WORK

Color images are represented in different color spaces such as RGB (Red Green Blue), HSV (Hue, Saturation, Value), YUV, YIQ, YCbCr (Luminance/Chrominance) etc. YCbCr is one of the best representations for steganography because the eye is sensitive to small changes in luminance but not in chrominance, so the chrominance part can be altered, without visually impairing the overall image quality much. Y is luminance component and Cb, Cr are the blue and red chrominance components respectively. The values in one color space can be easily converted into another color space using conversion formula [4]. Figure 2 shows the various components of Lena color image.

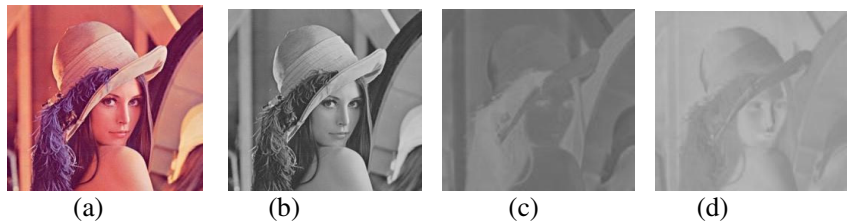

(a)      (b)      (c)      (d)

Figure. 2 (a) Lena, (b) luminance component Y of (a), (c) chrominance component Cb of (a), (d) chrominance component Cr of (a)

S. M. Masud Karim, et al., [5] proposed a new approach based on LSB using secret key. The secret key encrypts the hidden information and then it is stored into different position of LSB of image. This provides very good security. XIE Qing et al.,[6] proposed a method in which the information is hidden in all RGB planes based on HVS (Human Visual System). This degrades the quality of the stego image. In the method proposed by Sunny Sachdeva et al., [7] the Vector Quantization (VQ) table is used to hide the secret message which increases the capacity and also stego size. The method proposed by Rong-Jian Chen et al [8], presents the novel multi-bit bitwise adaptive embedding algorithm for data hiding by evaluating the most similar value to replace the original one. Sankar Roy et al., [9] proposed an improved steganography approach for hiding text messages within lossless RGB images which will suffer from withstanding the signal processing operations.

Minimum deviation of fidelity based data embedding technique has been proposed by J. K. Mandal et al, [10] where two bits per byte have been replaced by choosing the position randomly between LSB and up to fourth bit towards MSB. A DWT based frequency domain steganographic technique, termed as WTSIC is also proposed by the same authors, [11] where secret message/image bits stream are embedded in horizontal, vertical and diagonal components. Anjali Sejul, et al, [4] proposed an algorithm in which binary images are considered to be secret images which are embedded inside the cover image by taking the HSV (Hue, Saturation, Value) values of the cover image into consideration. The secret image is inserted into the cover image by cropping the cover image according to the skin tone detection and then applying the DWT. In this method the capacity is too low.

Saeed Sarreshtedari et al., [12] proposed a method to achieve a higher quality of the stego image using BPCS (Bit Plane Complexity Segmentation) in the wavelet domain. The capacity of each DWT block is estimated using the BPCS. Saddaf Rubab et al., [13] proposed a complex method using DWT and Blowfish encryption technique to hide text message in color image. In the paper by Kapre Bhagyashri et al, [14] a new singular value decomposition (SVD) and DWT based water mark technique is proposed in full frequency band in YUV color space. Nabin Ghoshal et al., uses a steganographic scheme for colour image authentication (SSCIA) [15] where the watermark image is embedded using DFT.

The proposed work is the modification of our previous work [16] in which the secret images are transmitted without actually embedding in the cover image. Only the keys are hidden in the cover image. The steps for embedding used in [16] are as follows:

- Represent the cover image C in YCbCr color space
- Obtain single level 2D DWT of secret-images S1, S2 and Cb, Cr component of C.
- The resulting transformed matrix consists of four sub-bands corresponding to LL, LH, HL and HH sub bands.
- LL sub band of Cb is used to hide one secret image and LL sub band of Cr is used to hide another secret image as follows:
    o The sub-images CLL and SLL are subdivided into non-overlapping blocks BCk1 (1 ≤ k1 < nc) and BSi (1 ≤ i < ns) of size 2x2 where nc, ns are the total number of non-overlapping blocks obtained from sub-images CLL and SLL respectively.
    o Every block BSi, is compared with block BCk1. The pair of blocks which have the least Root Mean Square Error is determined. A key is used to determine the address of the best matched block BCk1 for the block BSi.
    o Thus two keys K1 and K2 corresponding to two secret images S1 and S2 are obtained.
- The two keys are then encrypted using simple exclusive or operation with a key and run length encoded and then hidden in the cover image using Least Significant Bit technique. Blowfish technique can be used for encryption. The resultant image is a stego-image G.

The steps to extract the secret images from the stego image used in [16] are as follows:

- Represent the stego image G in YCbCr color space. Let it be GyGcbGcr
- Obtain the keys from Gcb and Gcr components
- Obtain DWT of Gcb and Gcr
- Divide GcbLL into non overlapping blocks of size 2x2
- Obtain the blocks that are nearest approximation to the original blocks of S1LL using K1
- Rearrange the blocks to obtain S1LLnew
- Obtain the secret image S1
- Similarly obtain S2 from GcrLL using K2

## 3. PROPOSED METHOD

In the proposed method, the cover is 256x256 color image. Two grey scale images of size 128 x128 are used as secret images. In this approach, the following steps are performed for encoding:

- Represent the cover image C in YCbCr color space
- Obtain single level IWT of secret-images S1, S2 and Cb, Cr component of C.
- The resulting transformed matrix consists of four sub-bands corresponding to LL, LH, HL and HH sub bands.

- LL sub band of Cb is used to hide one secret image and LL sub band of Cr is used to hide another secret image. Then the two keys K1 and K2 corresponding to two secret images are obtained using the same procedure used in our previous work [16] as described in section 2.
- The two keys are then encrypted using simple exclusive or operation with a key and run length encoded and then hidden in the cover image using IWT as follows:
    - Find the integer wavelet transform of Cb component of the cover image.
    - Replace the least significant bit planes of the higher frequency components of the transformed image by the bits of the key K1.
    - Obtain the inverse IWT of the resulting image to get the stego Cb component.
    - Similarly hide K2 in Cr component.
    - Represent the resultant image in RGB color space to obtain stego image G.

The secret images can now be extracted from the Cb and Cr components of the stego image as follows:

- Represent the stego image G in YCbCr color space. Let it be GyGcbGcr
- Obtain IWT of Gcb and Gcr and obtain the keys K1 and K2.
- Then secret images are obtained with the help of K1 and K2 following the same steps mentioned before for our previous work [16] in section 2.

## 4. EXPERIMENTAL RESULTS

The algorithm is tested in MATLAB. The wavelet tool box is used. The lifting wave *cdf 2.2* is used to find the integer wavelet transform. The results with different cover images and secret images are shown. Original cover and secret images are shown in Figure 3. Two cover images "baboon" and "peppers" (Figure 3(a) and 3(b)), each of size 256X256, are considered for testing the algorithm. The secret images considered are "earth", "football" and "moon" (Figure 3(c), 3(d), and 3(e)), each of size 128X128. The "football" and "earth" are embedded in "peppers". The resultant stego image is shown in Figure 4(a). The "earth" and "moon" are embedded in "baboon". The resultant stego image is shown in Figure 4(b). Extracted secret images from "peppers" are shown in Figure 4(c) and 4(d). Extracted secret images from "baboon" are shown in Figure 4(e) and 4(f). In all cases the average PSNR value of stego images is 44.7dB. The PSNR values of the extracted secret images are also approximately 44.7 dB. The PSNR values in dB in all cases for stego and extracted secret images are tabulated in Tables 1 and 2 respectively. Table 3 compares the PSNR value of the stego image in the proposed method and that in the other four methods. In all these the cover image considered is "peppers" and the secret images used are of comparable sizes. The average PSNR value in the proposed method is much higher than that in the other methods.

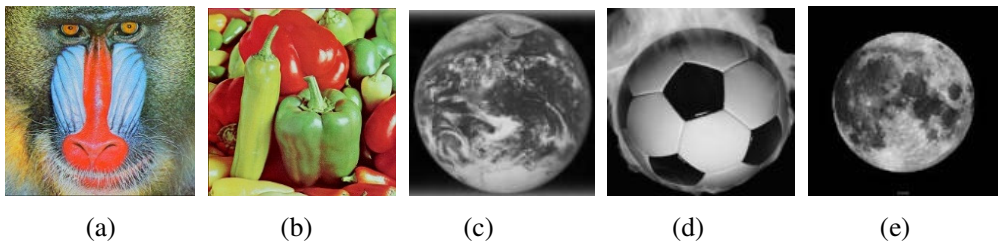

(a)　　　　(b)　　　　(c)　　　　(d)　　　　(e)

Figure 3 Cover and secret images:  (a) cover (baboon)   (b) cover (peppers)  (c) secret (earth)   (d) secret (football)   (e) secret (moon)

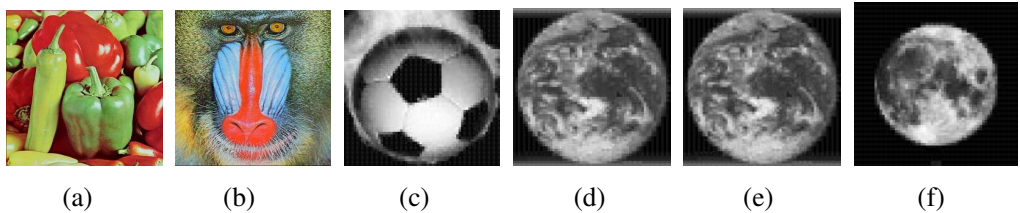

(a)   (b)   (c)   (d)   (e)   (f)

Figure 4 Stego and extracted secret images: (a) stego (football and earth as secret), (b) stego (moon and earth as secret), (c)-(f) extracted secret images: (c) football from peppers  (d) earth from peppers (e) earth from baboon   (f) moon from baboon

Table 1.  PSNR (in dB) of the stego image

| COVER IMAGE (256x256) | SECRET IMAGES (128x128) | PSNR |
|---|---|---|
| peppers | football and earth | 44.7 |
| baboon | earth and moon | 44.8 |

Table 2.  PSNR (in dB) of the extracted secret image

| COVER IMAGE (256x256) | SECRET IMAGES (128x128) | | |
|---|---|---|---|
| | football | earth | moon |
| peppers | 44.6 | 44.7 | |
| baboon | | 44.8 | 44.8 |

Table 3. Comparison of PSNR (in dB) of the stego image in different methods

| TECHNIQUE | PSNR |
|---|---|
| Mandal, J.K. et al. [10] | 39.6 |
| Mandal, J.K. et al. [11] | 42.4 |
| Kapre Bhagyashri, S. et al. [14] | 36.6 |
| Ghoshal, N. et al. [15] | 33.2 |
| PROPOSED | 44.7 |

## 5. CONCLUSIONS

In this paper, we observe that two secret images can be hidden in one color image and they can be regenerated without actually storing the image. This approach results in high quality of the stego-image having high PSNR values compared to other methods. However the disadvantage of the approach is that it is susceptible to noise if spatial domain techniques are used to hide the key. This can be improved if transform domain techniques are used to hide the key. The approach is very simple and the security level can be increased by using standard encryption techniques to encrypt the keys.